\documentclass{iopart}

\usepackage{setstack,iopams}
\usepackage{graphicx,subfigure}
\usepackage{cite}

\newcommand{\sign}{{\rm sign}}
\newcommand{\avg}[1]{\left\langle{#1}\right\rangle}
\newcommand{\ovl}[1]{\overline{#1}}
\newcommand{\ii}{{\rm i}}
\newcommand{\cavg}[1]{\left\langle\!\left\langle{#1}
	\right\rangle\!\right\rangle}
\newcommand{\erf}{{\rm erf}}
\newcommand{\pathavg}[1]{\left\langle{#1}\right\rangle_{{\rm paths}}}
\newcommand{\stavg}[1]{\left\langle{#1}\right\rangle_*}

\renewcommand{\l}{\left}
\renewcommand{\r}{\right}

\begin{document}

\title[Mixed majority-minority game
with random external information]{Statistical mechanics of the mixed
majority-minority game with random external information}

\author{A De Martino, I Giardina and G Mosetti}

\address{INFM--SMC and Dipartimento di Fisica, Universit\`a di Roma 
``La Sapienza'', P.le A. Moro 2, 00185 Roma, Italy}

\eads{\mailto{andrea.demartino@roma1.infn.it},
\mailto{irene.giardina@roma1.infn.it},
\mailto{giancarlo.mosetti@virgilio.it}}

\begin{abstract}
We study the asymptotic macroscopic properties of the mixed
majority-minority game, modeling a population in which two types of
heterogeneous adaptive agents, namely ``fundamentalists'' driven by
differentiation and ``trend-followers'' driven by imitation,
interact. The presence of a fraction $f$ of trend-followers is shown
to induce (a) a significant loss of informational efficiency with
respect to a pure minority game (in particular, an efficient,
unpredictable phase exists only for $f<1/2$), and (b) a catastrophic
increase of global fluctuations for $f>1/2$. We solve the model by
means of an approximate static (replica) theory and by a direct
dynamical (generating functional) technique. The two approaches
coincide and match numerical results convincingly.
\end{abstract} 

%%%%%%%%%%%%%%%%%%%%%%%%%%%%%%%%%%%%%%%%%%%%%%%%%%%%%%%%%%%%%%%%

\section{Introduction}

In recent years, a substantial amount of research has been focused on
model systems of heterogeneous adaptive agents interacting
competitively, as e.g. in games, markets or ecosystems, in the attempt
to understand the mechanisms by which real systems create exploitable
information, and to clarify the origin of their complex collective
behavior \cite{santafe}. The minority game, with its several variants,
is perhaps the most studied of such models \cite{webpage}. In its
simplest version, it describes a population of boundedly rational
players with fully heterogeneous beliefs who, at each round of the
game, make their strategic decisions basing on some public information
pattern (the ``state of the world'') aiming to be in the minority
group.  The minority-wins mechanism, which serves the purpose of
modeling competition for a scarce resource, translates into a strong
assumption on the behavioral traits and expectations of
players. Indeed, it turns out that in order to maximize their expected
utilities under the minority-wins rule, agents have to enhance their
initial heterogeneity and differentiate themselves as much as possible
from each other. This is rather intuitive: if agents would learn to
make decisions similarly to each other, being in a minority would
become a rather unlikely event. On the other hand, one might also
consider another tendency that is often encountered in real agents,
namely that toward imitation, say of an agent who believes that
his/her payoff is maximized when he/she acts according to the
majority. In this paper, we consider a mixed majority-minority game,
to study the effects of competition in a population formed by two
types of players, i.e. those whose short-term behavior is driven by
imitation (who play a majority game), and those who are instead
anti-imitative (and play a minority game).

From the viewpoint of economic modeling, our system represents a
simple abstraction for a market where two classes of economic agents,
namely ``fundamentalists'' and ``trend-followers'', interact. The
former -- see \cite{matteo,praga} for details -- create their
expectations under the assumption that the market price is close to
its ``fundamental'' value, i.e. to a stationary equilibrium, and
correspond to minority game players. The latter, instead, extrapolate
a trend from recent price increments and assume that the next
increment will occur in the direction of the trend (see also
\cite{gb2,sornette}); they correspond to majority game players. In
real markets, fundamentalists act as a kind of elastic force that
pulls the price toward its fundamental value, while trend-followers
destabilize the market by driving the price away from it. They are in
particular widely believed to be the main actors in the infamous buy
rushes known as ``bubbles''. Understandably then, modeling the
interplay between trend-followers and fundamentalists is a basic issue
in the theory of markets, and several models have been proposed (see
e.g. \cite{gb2,sornette,lm,gb} and references therein). In most cases,
however, insight can be gained only from numerical simulations due to
the complexity of the microscopic definitions. The mixed model we
consider here has the advantage of being simple enough to be
analytically tractable via the methods of statistical mechanics,
notwithstanding its phenomenological richness. The effects due to the
presence of trend-followers are fully discernible and an
interpretation in market terms is quite straightforward\footnote{For
another minority-game based market model with two different types of
agents, ``speculators'' and ``producers'', see \cite{cmz}.}. Besides,
the majority game is an interesting model in itself, that from the
theoretical viewpoint shares some features with the Hopfield model
\cite{hkp}. Surprisingly, it has not received much attention so far
\cite{decara,matmaj}.

This work is organized as follows. The basic definitions of the model
are given in Sec.~2, together with an outline of the results. The
static approximation to the analysis of the asymptotic macroscopic
properties is expounded in Sec.~3. It is based on the formal analogy
with zero-temperature spin glasses first derived in \cite{prl} for the
pure minority game, whose stationary states were shown to be
(approximately) given by the minima of a random Hamiltonian. In our
case, the resulting optimization problem is slightly more subtle and
its solution requires a negative dimensional replica theory of the
kind already used for ``minimax games'' \cite{varga}, close in spirit
to the method of partial annealing \cite{dfm}. Sec.~4 is devoted to
the dynamical solution of the ``batch'' version of the model, which is
carried out employing the generating functional technique \cite{dedo}
along the lines of \cite{hc,chs}. Some details about this calculation
are given in the Appendix. Finally, in Sec.~5, we formulate our
conclusions.

%%%%%%%%%%%%%%%%%%%%%%%%%%%%%%%%%%%%%%%%%%%%%%%%%%%%%%%%%%%%%%%%%

\section{Definitions and outline of the results}

The setup we consider is as follows. There are $N$ players and $P$
possible information patterns. For each player $i\in\{1,\ldots,N\}$
two strategies $\boldsymbol{a}_{ig}:\{1,\ldots,P\}\ni\mu\to
a_{ig}^\mu\in\{-1,+1\}$ are given ($g=1,2$) that map an information
pattern $\mu$ into a binary trading action $a_{ig}^\mu$
(`buy/sell'). (The generalization to $S$ strategies per agent is
possible but it is analytically less convenient.) We assume as usual
that $P$ scales with $N$ so that $P/N=\alpha$ remains finite in the
relevant limit $N\to\infty$ and that each $a_{ig}^\mu$ is selected
randomly with uniform probability in $\{-1,1\}$ at the beginning of
the game for all $i$, $\mu$ and $g$ and fixed. Strategies are
evaluated according to their ``performance'' $p_{ig}(n)$. At each
round $n$, players receive an information pattern $\mu(n)$ chosen at
random with uniform probability in $\{1,\ldots,P\}$
\cite{cava,rele}. Subsequently, each player picks his so-far
best-performing strategy, $\widetilde{g}_i(n)={\rm arg~max}_g
~p_{ig}(n)$, and formulates the bid it prescribes, i.e.
$a_{i\widetilde{g}_i(n)}^{\mu(n)}$.  The aggregate action of all
players at round $n$ (in economic terms, the ``excess demand'') is
just
\begin{equation}
A(n)=\frac{1}{\sqrt{N}}\sum_{i=1,N}a_{i\widetilde{g}_i(n)}^{\mu(n)}
\end{equation}
Once $A(n)$ is known, majority (resp. minority) game players reward
their strategies for which $a_{ig}^{\mu(n)}A(n)>0$
(resp. $a_{ig}^{\mu(n)}A(n)<0$). Hence the performance updating or
learning process takes place according to\footnote{We assume that
players ignore their market impact, i.e. that they behave as price
takers \cite{mcz}.}
\begin{equation}\label{pud}
p_{ig}(n+1)-p_{ig}(n)=\epsilon_i a_{ig}^{\mu(n)}A(n)\qquad(g=1,2)
\end{equation}
where $\epsilon_i=-1$ for minority game players and $\epsilon_i=+1$
for majority game players, and the game moves into the next round. The
$\epsilon_i$'s can be seen as an additional family of quenched r.v.'s
(besides the $a_{ig}^\mu$'s) with probability density
$P(\epsilon_i)=f\delta_{\epsilon_i,+1}+(1-f)\delta_{\epsilon_i,-1}$.

For later use, it is convenient to introduce the ``preferences''
$y_i(n)=(p_{i1}(n)-p_{i2}(n))/2$ and the quantities
$\xi_i^\mu=(a_{i1}^\mu-a_{i2}^\mu)/2$,
$\omega_i^\mu=(a_{i1}^\mu+a_{i2}^\mu)/2$ and
$\Omega^\mu=N^{-1/2}\sum_{i=1,N}\omega_i^\mu$, using which
(\ref{pud}) can be recast as an equation for $y_i(n)$:
\begin{equation}\label{dyna}
y_i(n+1)-y_i(n)=\epsilon_i \xi_i^{\mu(n)}
[\Omega^{\mu(n)}+\frac{1}{\sqrt{N}}\sum_{j=1,N}\xi_j^{\mu(n)}s_j(n)]
\end{equation}
where $s_i(n)=\sign[y_i(n)]$. When $y_i(n)>0$ (resp. $y_i(n)<0$) agent
$i$ selects strategy $g=1$ (resp. $g=2$) and $s_i(n)=+1$
(resp. $s_i(n)=-1$). As in the pure minority game, this stochastic
(indeed, Markovian) dynamics is a zero-temperature process that
violates detailed balance so that, strictly speaking, no equilibrium
state exists. 

As usual, one is interested in characterizing the macroscopic
properties of the stationary state (if any exists) of
(\ref{dyna}). Two quantities have been introduced to this aim. As a
measure of global efficiency one uses the ``volatility''
\begin{equation}\label{s2}
\sigma^2=\avg{A^2}=\lim_{T\to\infty}\frac{1}{T-T_{{\rm
eq}}}\sum_{n=T_{{\rm eq}},T} A(n)^2
\end{equation}
that is, the magnitude of market fluctuations ($\avg{A}=0$ by
construction). Intuitively, efficiency is higher the smaller is
$\sigma^2$. As a reference value, it is reasonable to take
$\sigma^2=1$, which corresponds to ``random players'' who at each
round randomize uniformly between the two possible actions. When
$\sigma^2<1$ one can say that agents are, to some degree,
cooperating. From the viewpoint of information creation, the relevant
quantity is instead the ``predictability'' or ``available
information''
\begin{equation}\label{H}
\fl H=\frac{1}{P}\sum_{\mu=1,P}\avg{A|\mu}^2\qquad{\rm with}~~
\avg{A|\nu}=\lim_{T\to\infty}\frac{1}{T-T_{eq}}\sum_{n=T_{eq},T}
A(n)\delta_{\mu(n),\nu}
\end{equation}
whose meaning is discussed at length in the literature (see
e.g. \cite{mcz,cm}). The idea is that when $H>0$ there exists at least
one state of the world, say $\mu$, such that $\avg{A|\mu}\neq 0$,
i.e. for which there is an action that is more likely to be the
winning action. An external agent entering the game could hence
exploit this information to have a gain. The fact that $H>0$ signals
an inefficiency of the market. Regimes with $H>0$ are dubbed
`asymmetric', at odds with `symmetric' ones with $H=0$ where the
game's outcome is not predictable.

\begin{figure}[!]
\begin{center}
\subfigure{\scalebox{.41}{\includegraphics{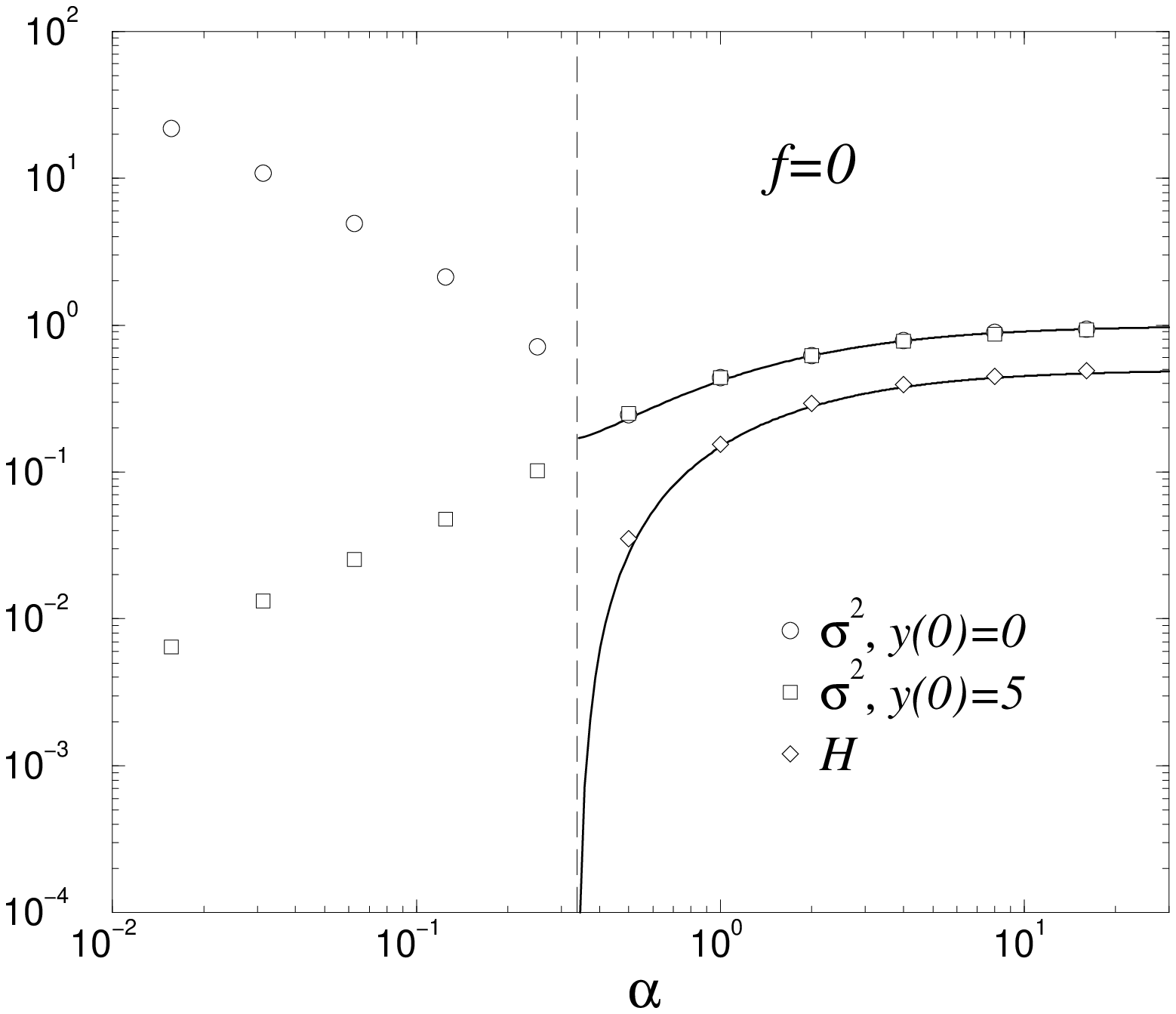}}}
\subfigure{\scalebox{.41}{\includegraphics{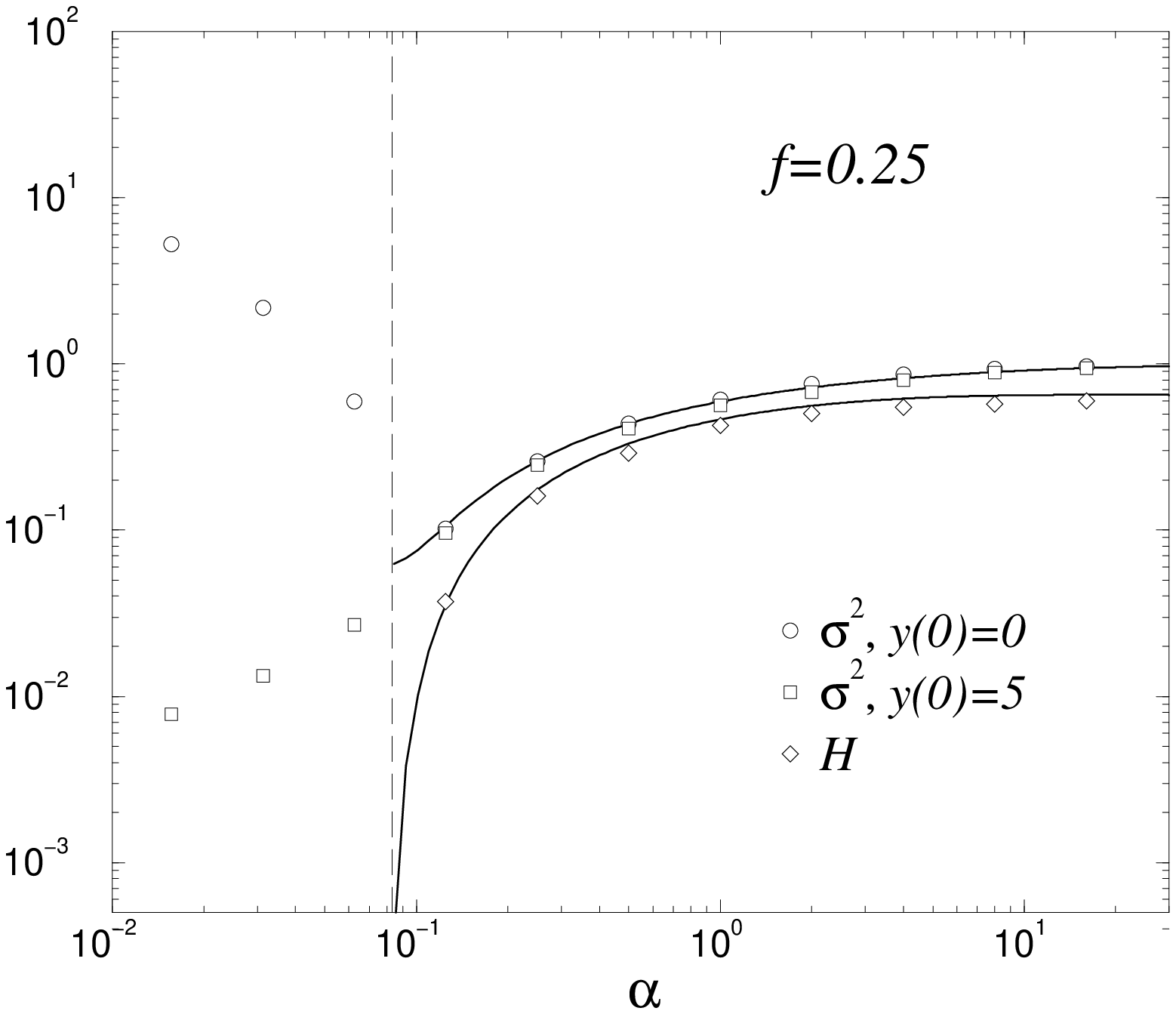}}}
\subfigure{\scalebox{.41}{\includegraphics{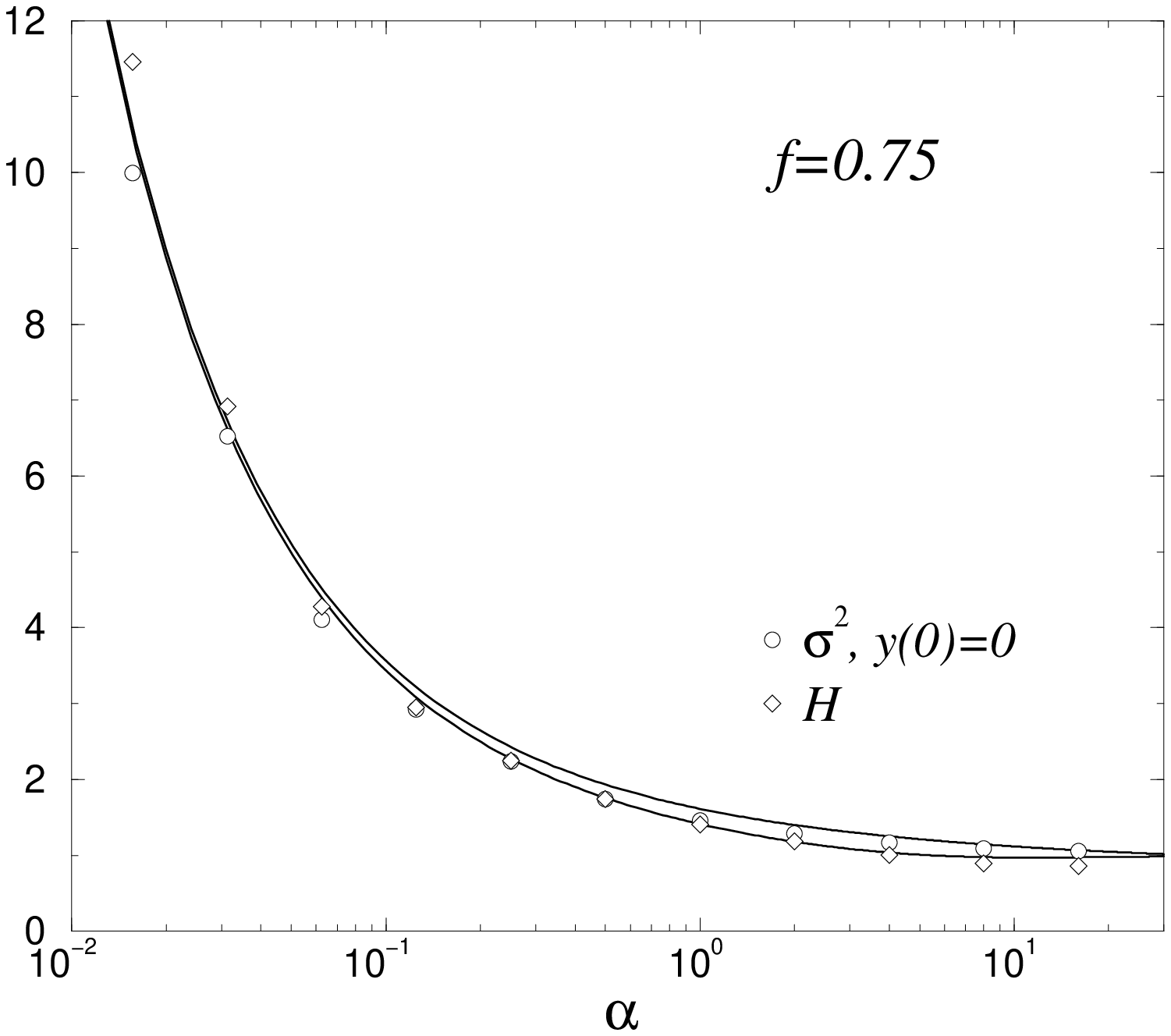}}}
\subfigure{\scalebox{.41}{\includegraphics{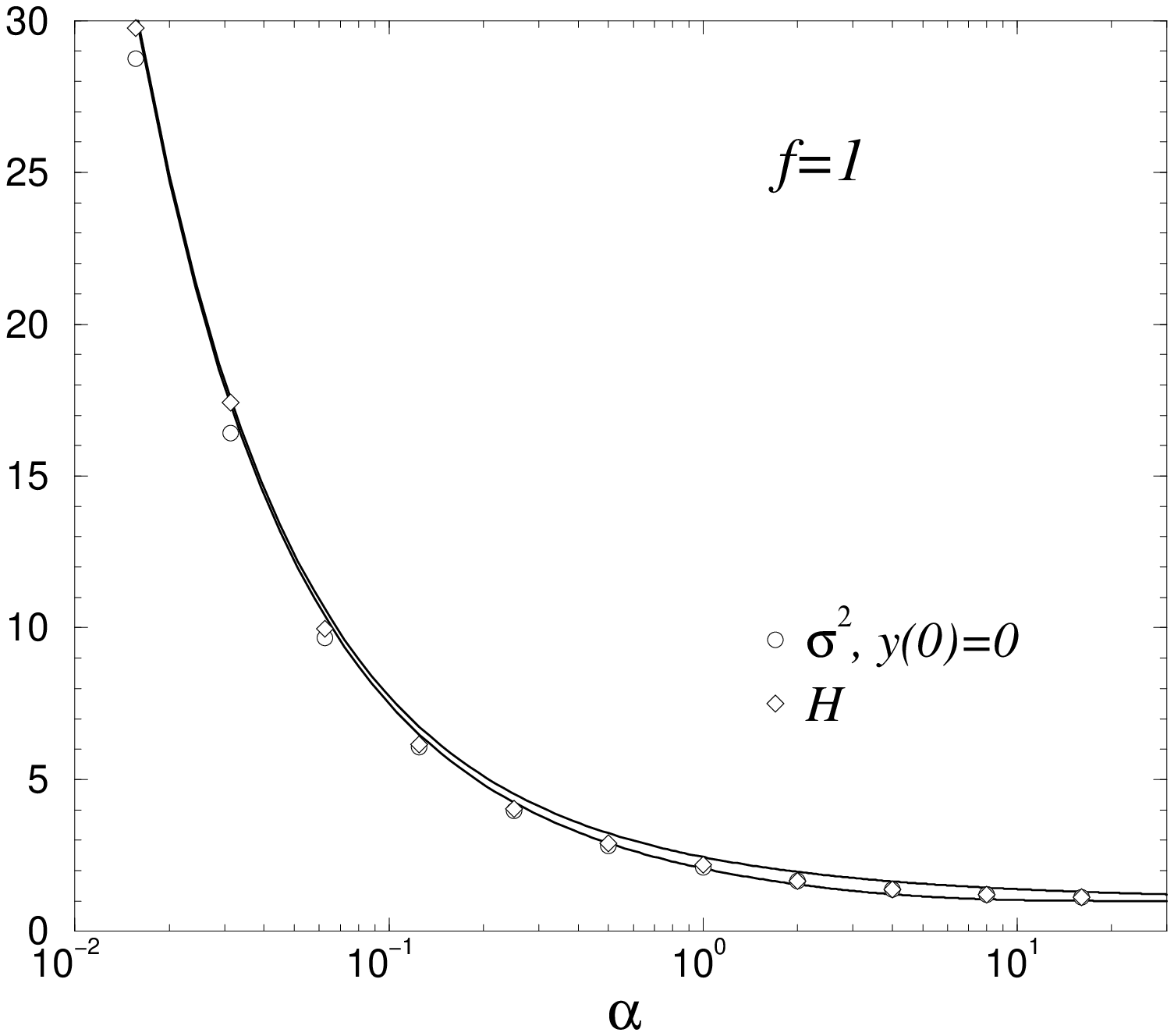}}}
\caption{\label{figs2H}Behavior of $\sigma^2$ and $H$ vs $\alpha$ for
$f=0,0.25,0.75,1$. Markers represent results from numerical
simulations with homogeneous initial conditions, averaged over 200
disorder samples. The dashed vertical lines give the location of
$\alpha_c$ (from theory). Continuous lines represent analytical
approximations (valid only for $\alpha>\alpha_c$). Results for $H$ are
compared with the static approximation of Sec. 3, while those for
$\sigma^2$ are compared with the dynamical results of Sec. 4. The
logarithmic scale on the $y$-axis in the upper panels has been used to
stress the dependence of $\sigma^2$ on the initial conditions for
$\alpha<\alpha_c$. In the lower panels, the upper curves correspond in
both figures to the static results for $H$.}
\end{center}
\end{figure}

In the limit $N\to\infty$, $\sigma^2$ and $H$ depend on $\alpha$ (as
in the pure minority game) and $f$. Computer simulations of
(\ref{dyna}) suggest the following scenario (see Fig.~\ref{figs2H}).
For $f<1/2$, a minority-game type of behavior is recovered, with an
asymmetric phase ($H>0$) at high $\alpha$ separated by a symmetric one
($H=0$) at low $\alpha$. The transition point\footnote{We remind that
strictly speaking this is not an equilibrium phase transition since
(\ref{dyna}) violates detailed balance.} $\alpha_c$ decreases as $f$
increases, hence the symmetric phase shrinks as more and more
trend-followers join the game, indicating that they provide an
additional exploitable `signal'. Market fluctuations tend to the
random limit $\sigma^2=1$ for large $\alpha$ and decrease with
$\alpha$ until the critical point is reached. In the sub-critical
phase, the stationary state depends strongly on the initial conditions
of (\ref{dyna}), and both high-volatility and low-volatility states
can be reached starting from slightly different
configurations\footnote{Clearly, if the initial conditions $y_i(0)$
contain a sufficiently large bias toward one of the strategies, all
players will always use the same strategy, which will evidently result
in the `random trading' state with $\sigma^2=1$.}. For $f>1/2$,
instead, trend-followers dominate the game and the global efficiency
decreases steadily with $\alpha$ and $f$. The market is asymmetric
($H>0$) for all $\alpha$ and the difference between $\sigma^2$ and $H$
diminishes as $f$ increases. For $f=1$, one has $\sigma^2=H$. The
dependence of $\sigma^2$ on the initial conditions is arguably very
weak (obviously provided initial conditions are not too biased). The
case $f=1/2$ possesses some special features and will be treated
separately \cite{prep}.

In order to get some theoretical insight, one can follow the line of
reasoning adopted for the pure minority game, for which it was shown
by constructing the continuous-time limit of (\ref{dyna}) that the
average asymptotic value of $s_i$, denoted by $m_i$, can be obtained
by minimizing the random function
\begin{equation}\label{ham}
\mathcal{H}(\boldsymbol{m})=\frac{N}{P}\sum_{\mu=1,P}
[\Omega^\mu+\frac{1}{\sqrt{N}}\sum_{i=1,N}\xi_i^\mu
m_i]^2
\end{equation}
where $\boldsymbol{m}=\{m_i\}$. (Notice that the $m_i$'s are `soft'
spin: $-1\leq m_i\leq 1$.) We will not discuss here the limitations of
this approximation and refer the reader to the original literature
\cite{prl,cggs,gms,com,repl,mc,ch} for a critical discussion.  In the
limit $N\to\infty$, this problem could be tackled using spin-glass
techniques, because
\begin{equation}\label{min}
\lim_{N\to\infty}\min_{\boldsymbol{m}}~
\frac{\mathcal{H}(\boldsymbol{m})}{N}=-\lim_{\beta\to\infty}
\lim_{N\to\infty}\frac{1}{\beta N}~\ovl{\log Z(\beta)}
\end{equation}
(here, $Z(\beta)=\int e^{-\beta\mathcal{H}}d\boldsymbol{m}$ and the
over-line denotes an average over disorder). The evaluation of
$\ovl{\log Z}$ requires the replica trick \cite{mpv}. For
$\alpha>\alpha_c$, $\mathcal{H}$ has a unique minimum, hence the
stationary state can be fully described by the replica-symmetric
solution of (\ref{min}).

This argument can be easily reformulated for the pure majority
game. The corresponding optimization problem turns out to be
\begin{equation}\label{maj}
\max_{\boldsymbol{m}}~ \mathcal{H}(\boldsymbol{m})\quad{\rm ~or,
equivalently,}\quad\min_{\boldsymbol{m}}~ -\mathcal{H}(\boldsymbol{m})
\end{equation}
A few comments are in order. First, it is easy to see that
$H=\mathcal{H}/N$, which implies that minority game players roughly
tend to minimize the available information, while majority ones tend
to maximize it. Second, at odds with $\mathcal{H}$, $-\mathcal{H}$
possesses many minima, hence the stationary state of the majority game
will always depend on the initial conditions of the dynamics (even
though the macroscopic observables $\sigma^2$ and $H$ might take on
the same or very similar values in all minima). Basing on well-known
properties of the Hopfield model \cite{hkp}, one expects the true
minima of $-\mathcal{H}$ to be described by solutions of (\ref{maj})
that break replica symmetry. Moreover, as happens in attractor neural
networks with extensively many patterns, a ``retrieval'' phase is to
be expected for small enough $\alpha$ where, due to correlations
between the initial conditions and one specific pattern, say $\mu=1$,
the overlap $o^\mu(\boldsymbol{m})=N^{-1/2}\sum_{i=1,N}\xi_i^\mu m_i$
is $\Or(N^{-1/2})$, and vanishing as $N\to\infty$, for all $\mu$'s
except $\mu=1$, for which it is finite. The fact that agents can
`condense' around a given pattern implies that every time that pattern
is presented to them a buy (or sell) rush takes place. Solving
(\ref{maj}) is hence a non-trivial task in itself, and requires a
detailed study \cite{matmaj}.

Generalizing to our case, one finds that the stationary $m_i$'s for
the mixed majority-minority model can be obtained by solving the
following problem:
\begin{equation}
\max_{\boldsymbol{m}_2}~\min_{\boldsymbol{m}_1}
~\mathcal{H}(\boldsymbol{m}_1,\boldsymbol{m}_2)
\end{equation}
where $\boldsymbol{m}_1$ (resp. $\boldsymbol{m}_2$) denote collective
the $m_i$ variables of minority (resp. majority) game
players\footnote{Notice that the $\min$ and $\max$ operations can be
interchanged. In general, this leads to different solutions. In our
case, however, one can verify that the main results would not change,
though the intermediate steps (e.g. the definition of $\gamma$) would
vary}. Hence the mixed game where both minority and majority players
are present at the same time requires a minimization of $\mathcal{H}$
in certain directions (the minority ones) and a maximization in others
(the majority ones). Again, this problem can be tackled by a replica
theory. The idea \cite{varga} is to introduce two `inverse
temperatures' $\beta_1$ and $\beta_2$ for minority and majority
players respectively, such that
\begin{equation}\label{minimax}
\max_{\boldsymbol{m}_2}~\min_{\boldsymbol{m}_1}
~\mathcal{H}(\boldsymbol{m}_1,\boldsymbol{m}_2)=
\lim_{\beta_1,\beta_2\to\infty}
\frac{1}{\beta_2}~\ovl{\log\mathcal{Z}(\beta_1,\beta_2)}
\end{equation}
with the following generalized partition function:
\begin{equation}\label{zeta}
\fl \mathcal{Z}(\beta_1,\beta_2)=\int d\boldsymbol{m}_2 ~
e^{\beta_2\l[-\frac{1}{\beta_1}\log\int d\boldsymbol{m}_1~
e^{-\beta_1\mathcal{H}}\r]}=\int d\boldsymbol{m}_2\l[\int
d\boldsymbol{m}_1 ~e^{-\beta_1\mathcal{H}}\r]^{-\gamma}
\end{equation}
where $\gamma=\beta_2/\beta_1>0$. In physical jargon, this describes a
system where: first, the $\boldsymbol{m}_1$ variables are thermalized
at a positive temperature $1/\beta_1$ with Hamiltonian $\mathcal{H}$
at fixed $\boldsymbol{m}_2$; then, the $\boldsymbol{m}_2$ variables
are thermalized at a negative temperature $-1/\beta_2$ with an
effective Hamiltonian $\mathcal{H}_{{\rm eff}}$ defined by
$-\beta_1\mathcal{H}_{{\rm eff}}(\boldsymbol{m}_2)=\log\int
d\boldsymbol{m}_1~ e^{-\beta_1 \mathcal{H}}$. The disorder average can
be carried out with the help of a `nested' replica trick. First, one
replicates the minority variables by treating the exponent $-\gamma$
as a positive integer $R$ (in the end, the limit $R\to-\gamma<0$ must
be taken). (\ref{zeta}) thus becomes
\begin{equation} 
\fl \mathcal{Z}=\int d\boldsymbol{m}_2\l[\int d\boldsymbol{m}_1
~e^{-\beta_1\mathcal{H}}\r]^{R}=\int d\boldsymbol{m}_2
\l[\int e^{-\beta_1\sum_{r}\mathcal{H}(\{\boldsymbol{m}_1^r\},
\boldsymbol{m}_2)}\prod_{r=1,R}d\boldsymbol{m}_1^r\r]
\end{equation} 
Then a second replication is needed\footnote{We remind the reader that
replica theories use the fact that $\ovl{\log Z}=\lim_{n\to
0}(1/n)\log~\ovl{Z^n}$.}, this time on the $\boldsymbol{m}_2$
variables:
\begin{equation} 
\mathcal{Z}^n=\int e^{-\beta_1\sum_{a,r}
\mathcal{H}(\{\boldsymbol{m}_1^{ar}\},\{\boldsymbol{m}_2^a\})}
\prod_{a=1,n}\prod_{r=1,R}d\boldsymbol{m}_1^{ar}
d\boldsymbol{m}_2^a
\label{zn}
\end{equation} 
At this point we have two replica indexes with different roles: the
$a$ replicas have been introduced to deal with the disorder, and their
number $n$ will eventually go to zero, as usual; the $r$ replicas have
been introduced to deal with the negative temperature, and their
number $R$ must be set to a negative value\footnote{This kind of limit
is not completely new in replica theories; this is what is usually
done for example to express determinants via a bosonic integral
representation, see for instance \cite{kur} for a discussion and
\cite{cgp} for an application.}. Majority variables bear just one
index, while minority ones have two. We can interpret this fact by
saying that $\boldsymbol{m}_2^a$ indicates a particular configuration
of the majority variables, i.e. a given manifold in the whole
$\boldsymbol{m}$ space; and $\boldsymbol{m}_1^{ar}$ indicates the
minority coordinates in that particular manifold.

In Sec.~3 we will solve (\ref{minimax}) in the limit $N\to\infty$
using (\ref{zn}) as a starting point. Evidently, retrieval solutions
for the majority part become increasingly important as $f$ gets
bigger. We will however neglect this aspect (which in the mixed case
leads to a serious lengthening). Results obtained in this way give a
very good agreement with numerical simulations, suggesting that
retrieval doesn't substantially affect the average macroscopic
properties of the game. Of course, it is expected to play a very
important role for phenomena that are local in time (like
``bubbles''). Besides this static approximation, we will also tackle
the dynamics (\ref{dyna}) straightforwardly, resorting to the
generating-functional method to carry out the disorder-average. Again,
we will neglect the possibility of retrieval. Following \cite{hc}, we
will focus on the `batch' version of the model. Dynamical results
obtained in this way turn out to coincide nicely with their static
counterpart and suggest that the transition occurring at $\alpha_c$
for $f<1/2$ is related essentially to the onset of anomalous response,
as in the pure minority game. We will calculate the critical line
$\alpha_c(f)$, showing that $\alpha_c\downarrow 0$ as $f\uparrow
1/2$. For $f>1/2$, the response is always finite and the macroscopic
properties are dominated by the contribution of trend followers.

%%%%%%%%%%%%%%%%%%%%%%%%%%%%%%%%%%%%%%%%%%%%%%%%%%%%%%%%%%%%%%%%%

\section{Statics}

To begin, let us re-write the Hamiltonian (\ref{ham}) as
\begin{equation}
\fl \mathcal{H}(\boldsymbol{m}_1,\boldsymbol{m}_2)=
\frac{1}{P}\sum_{\mu=1,P}\l[\sum_{i=1,N}\omega_i^\mu +\sum_{j\in
N_1}\xi_j^\mu m_{1j}+\sum_{k\in N_2}\xi_k^\mu m_{2k}\r]^2
\end{equation}
where $N_1$ (resp. $N_2$) denotes both the set and the cardinality of
the set of minority (resp. majority) game players. The replicated
Hamiltonian entering (\ref{zn}) is
\begin{equation}
\fl \mathcal{H}(\{\boldsymbol{m}_1^{ar}\},\{\boldsymbol{m}_2^a\})=
\frac{1}{P}\sum_{\mu=1,P}\l[\sum_{i=1,N}\omega_i^\mu +\sum_{j\in
N_1}\xi_j^\mu m_{1j}^{ar}+\sum_{k\in N_2}\xi_k^\mu m_{2k}^a\r]^2
\end{equation}
We can as usual linearize the exponential in (\ref{zn}) via a
Hubbard-Stratonovich transformation introducing some auxiliary
Gaussian variables $z_{ar}^\mu$. Subsequently, the average over the
disorder can be easily performed using the distribution $P(a_{ig}^\mu)
= 1/2 (\delta_{a_{ig}^\mu,1}+\delta_{a_{ig}^\mu,-1})$ ($g=1,2$) and
the definitions of $\omega_i^\mu$ and $\xi_i^\mu$. One obtains
\begin{eqnarray}\label{intermetzo}	
\fl \overline{\mathcal{Z}^n}=\int [\prod_{a,r}d\boldsymbol{m}_1^{ar}
d\boldsymbol{m}_2^a][\prod_{\mu,a,r}\frac{dz_{ar}^\mu}{\sqrt{2\pi}}] ~
e^{-\sum_{\mu}\sum_{ar}\frac{(z_{ar}^\mu)^2}{2}}\times\\ \times
e^{-\frac{\beta_1}{2\alpha}\sum_\mu \sum_{abrs} z_{ar}^\mu
z_{bs}^\mu(1+(1-f) \frac{1}{N_1}\sum_{j \in N_1}
m_{1j}^{ar}m_{1j}^{bs}+f\frac{1}{N_2} \sum_{k \in N_2}m_{2k}^a
m_{2k}^b)}\nonumber
\end{eqnarray}
It is now convenient to define the overlaps
\begin{equation}
Q_{ar,bs}= \frac{1}{N_1}\sum_{j\in N_1} m_{1j}^{ar}m_{1j}^{bs}
\quad{\rm and}\quad P_{ab}=\frac{1}{N_2}\sum_{k\in N_2}
m_{2k}^{a}m_{2k}^{b}
\end{equation}
inserting them in (\ref{intermetzo}) via $\delta$-distributions
with Lagrange multipliers $\widehat{Q}_{ab,rs}$ and
$\widehat{P}_{ab}$. Notice that the overlap matrices $\mathsf{Q}$ and
$\mathsf{P}$ are $nR$-dimensional and $n$-dimensional,
respectively. In this way the site dependence can be easily dealt
with, so that after a little algebra one gets (all numerical factors
are `hidden' in the $D(\cdot,\cdot)$ shorthand):
\begin{equation}
\overline{\mathcal{Z}^n}=\int e^{NS(\mathsf{Q},
\mathsf{\widehat{Q}},\mathsf{P},\mathsf{\widehat{P}})}~
D(\mathsf{Q},\mathsf{\widehat{Q}})
D(\mathsf{P},\mathsf{\widehat{P}})
\label{inte}
\end{equation}
where the effective action $S$ is given by ($a,b=1,\ldots,nR~$;
$~r,s=1,\ldots, R$)
\begin{eqnarray}\label{action}
\fl
S(\mathsf{Q},\mathsf{\widehat{Q}},\mathsf{P},\mathsf{\widehat{P}})=
-\frac{\alpha}{2}\log\det\mathsf{T}-\ii~[(1-f)\Tr(\mathsf{\widehat{Q}}
\mathsf{Q})+ f \Tr(\mathsf{\widehat{P}}\mathsf{P})]+\nonumber\\\fl
+(1-f) \log\int_{-1}^{+1}[\prod_{a,r}dm_1^{a,r}]
~e^{\ii\sum_{abrs}m_1^{ar}\widehat{Q}_{ar,bs}m_1^{bs}}
+f\log\int_{-1}^{+1} [\prod_a
dm_2^a]~e^{\ii\sum_{ab}m_2^a\widehat{P}_{ab}m_2^b}\nonumber
\end{eqnarray}
with
\begin{equation}
\mathsf{T}=\mathsf{I}_{nR}+\frac{\beta_1}{\alpha}[\mathsf{E}_{nR}
+(1-f)\mathsf{Q} + f \mathsf{P}\otimes\mathsf{E}_R]
\label{matricet}
\end{equation}
$\mathsf{I}_K$ stands for the $K\times K$ identity matrix while
$\mathsf{E}_K$ denotes the $K\times K$ matrix with all elements equal
to $1$. $\otimes$ is the Kronecker product. In (\ref{action}) one can
easily recognize some parts coming from the minority agents (those
proportional to $(1-f)$) and others coming from the majority
agents. These contributions are {\it not} factorized (in that event,
the mixed problem would be trivial) but are interconnected via the
determinant of $\mathsf{T}$.

To proceed further, one has to formulate Ans\"atze for the overlap
matrices and then perform the integral (\ref{inte}) in the limit
$N\to\infty$ by the steepest descent method. Let us first arrange
$\mathsf{Q}$ in a convenient matrix form. We choose to order the
indexes in such a way that each row is characterized by a couple
$(a,r)$; along the row, the index $a$ is first kept fixed while $r$
varies from $1$ to $R$. $\mathsf{Q}$ is thus naturally subdivided in
blocks of size $R\times R$, the blocks along the diagonal
corresponding to a given value of $a=b$. We remind that keeping $a$
fixed corresponds to selecting, in the global configuration space, a
well defined manifold with $\boldsymbol{m}_2=\boldsymbol{m}_{2}^a$
inside which $\mathcal{H}$ is minimized with respect to the
$\boldsymbol{m}_1$ variables. $Q_{ar,as}$ can be thus interpreted as
the overlap between two configurations of the same constrained
minority problem. It is natural to assume for these diagonal
sub-blocks the same matrix structure of a pure minority game, that is
a symmetric form with a diagonal element $Q$ and an off-diagonal one
$q_1$. On the other hand, elements of the type $Q_{ar,bs}$ with $a\neq
b$ correspond to overlaps between two minority configurations in
different majority manifolds, and the simplest choice one can make is
to take $Q_{ar,bs}=q_0$ for all of them. In this way $\mathsf{Q}$
assumes what is called a 1-step RSB (replica symmetry broken) form
\cite{mpv}:
\begin{equation}\label{qu}
Q_{ar,bs}=(Q-q_1) \delta_{ab}\delta_{rs} + (q_1-q_0) \epsilon_{arbs}+
q_0
\end{equation}
where the tensor $\epsilon_{arbs}$ is equal to $1$ in the diagonal $R
\times R$ blocks with $a=b$, and $0$ elsewhere. Notice that, contrary
to standard replica calculations, here the block size $R$ is not a
variational parameter, but its value is fixed by the nature of the
problem. For consistency, we adopt the same Ansatz for the conjugated
matrix $\widehat{\mathsf{Q}}$. The choice for the $n\times n$ matrices
$\mathsf{P}$ and $\widehat{\mathsf{P}}$ is on the other hand more
straightforward: we will consider the simple replica-symmetric Ansatz
\begin{equation} \label{ppi}
P_{ab}=(P-p_0)\delta_{ab}+p_0
\end{equation} 
and take an analogous form for $\widehat{\mathsf{P}}$.

Putting (\ref{qu}) and (\ref{ppi}) into (\ref{action}), and using the
conventional re-scalings $\mathsf{\widehat
{Q}}=(-i\beta_1^2\alpha/2)\mathsf{\Omega}$ and $\mathsf{\widehat{P}}=
(-i\beta_1^2\alpha/2)\mathsf{G}$, the `free energy' density
$F=-S/(\beta_1 n)$ turns out to be given, in the limit $n\to 0$, by
\begin{eqnarray} \label{effe}
\fl F=\frac{\alpha
R}{2\beta_1}\log\l[1+(1-f)\frac{\beta_1}{\alpha}(Q-q_1)\r]+
\frac{\beta_1 R\alpha(1-f)}{2}\l[\Omega Q+(R-1)\omega_1 q_1-R\omega_0
q_0\r]+\nonumber\\ \fl +\frac{\alpha}{2
\beta_1}\log\l[1+R\beta_1\frac{(1-f) (q_1-q_0)+f \
(P-p_0)}{\alpha+(1-f)\beta_1 (Q-q_1)}\r]+\frac{\beta_1\alpha}{2} f
(GP-g_0p_0)+\nonumber \\\fl +\frac{\alpha R}{2}\frac{1+(1-f) q_0+f
p_0}{\left[\alpha +(1-f) \beta_1(Q-q_1)\right] \left[R\beta_1(1-f)
(q_1-q_0)+f (P-p_0)\right]}+\\ \fl -\frac{1-f}{\beta_1}\int dz\mathcal
P(z)\log\int dy\mathcal P(y)\l[\int_{-1}^1dm_1~e^{-\beta_1
V_{zy}(m_1)}\r]^R+\nonumber\\ \fl -\frac{f}{\beta_1}\int dz\mathcal
P(z)\log\int_{-1}^1 dm_2 ~e^{-\beta_1 V_z(m_2)} \nonumber
\end{eqnarray} 
where $\mathcal{P}(x)=e^{-x^2/2}/\sqrt{2\pi}$ and
\begin{eqnarray}
V_{zy}(m_1)=-z\sqrt{\alpha \omega_0}y
m_1\sqrt{\alpha(\omega_1-\omega_0)}- \frac{\alpha
\beta_1}{2}(\Omega-\omega_1)m_1^2 \\ V_z(m_2)=-\sqrt{\alpha
g_0}zm_2-\frac{\alpha\beta_1}{2}(G-g_0)m_2^2
\end{eqnarray}

The replica recipe now prescribes an extremization of (\ref{effe})
with respect to its ten variational parameters (namely $Q$, $q_0$,
$q_1$, $P$, $p_0$ and their conjugate variables), because when
$N\to\infty$ it is easy to see that
\begin{equation}\label{mimma}
\lim_{N\to\infty}\max_{\boldsymbol{m}_2}~\min_{\boldsymbol{m}_1}
~\frac{\mathcal{H}}{N}=\lim_{\beta_1,\beta_2\to\infty} \frac{F({\rm
saddle~ point})}{R}
\end{equation}
This leaves us with a set of ten equations in ten variables. Defining
\begin{eqnarray}
\chi_1=\frac{\beta_1}{\alpha}(Q-q_1)-\frac{\beta_2}{\alpha}(q_1-q_0)\\
\chi_2=\frac{\beta_2}{\alpha}(P-p_0)\\ \chi=(1-f)\chi_1-f\chi_2
\label{susc}
\end{eqnarray}
and using the shorthands
\begin{equation}
\fl \cavg{\cdots}=\int dz \mathcal P(z)\left[\frac{\int dy
\mathcal{P}(y)\left[\mathcal{Q}^{R-1}\int_{-1}^{1}dm_1~\cdots~e^{-\beta_1
V_{yz}(m_1)}\right]}{\int dy \mathcal
P(y)\left[\mathcal{Q}^R\right]}\right]
\label{ave1}
\end{equation} 
$\mathcal{Q}=\int_{-1}^{1}dm_1~e^{-\beta_1 V_{yz}(m_1)}$ being a
 normalization integral, and
\begin{equation}
\avg{\cdots}_2=\frac{\int_{-1}^{1}dm_2~\cdots~ e^{-\beta_1 V_z(m_2)}}
{\int_{-1}^{1}dm_2~e^{-\beta_1 V_z(m_2)}}
\label{ave2}
\end{equation}
we find the following system:
\begin{eqnarray}
Q=\cavg{m_1^2} \label{prima}\\ \beta_1 R q_1+\beta_1(Q-q_1)=
\frac{\cavg{ym_1}}{\sqrt{\alpha(\omega_1-\omega_0)}}\\
\alpha\chi_1=\frac{\cavg{zm_1}}{\sqrt{\alpha\omega_0}}\\
\beta_1(\Omega-\omega_1)=-\frac{1}{\alpha+\beta_1(1-f)(Q-q_1)}\\
\omega_1-\omega_0=\frac{(1-f)(q_1-q_0)+ f
(P-p_0)}{\alpha(1+\chi)[\alpha+\beta_1(1-f)(Q-q_1)]}\\
\omega_0=\frac{1+(1-f) q_0+fp_0}{\alpha^2(1+\chi)^2}\label{majo1}\\
P=\avg{m_2^2}_2\\ g_0=R^2\omega_0\label{majo2}\\
\alpha\chi_2=-R\frac{\avg{zm_2}_2}{\sqrt{\alpha g_0}}\label{majo3}\\
\beta_1(G-g_0)=-\frac{R}{\alpha(1+\chi)}\label{ultima}
\end{eqnarray}

Some observations about these equations are in order. First, if we set
$f=0$ we recover exactly the saddle point equations for a pure
minority game problem at inverse temperature $\beta_1$. For what
concerns the $\chi$'s, it will soon become clear that $\chi_1$ is the
susceptibility of minority agents and, when $f=0$, it reproduces the
susceptibility of a pure minority game, while $\chi_2$ is the
susceptibility of majority agents. On the other hand, $\chi$ is
evidently {\em not} the global susceptibility. This is a consequence
of the fact that to treat minority and majority players within the
same formalism we had to introduce the effective negative inverse
temperature $-\beta_2$.

Solving the above system at finite temperature(s) is a quite difficult
task. Fortunately, in this case we are only interested in the limit of
zero temperature(s), in which the solution of
(\ref{prima}--\ref{ultima}) turn out not to depend explicitly on $R$,
provided $G$ and $g_0$ are rescaled by $R^2$. Specifically, we look
for solutions with $q_0\to q_1\to Q$ and $p_0\to P=1$ such that
$\chi_1$, $\chi_2$ and $\chi$ remain finite. These assumptions are
justified for minority variables by the existence of just one global
minimum of $\mathcal{H}$ (which also means that the minimum is unique
in each manifold with given $\boldsymbol{m}_2$). On the other hand,
they are more questionable for majority variables, since the maxima of
$\mathcal{H}$ are numerous and disconnected (they occur evidently in
the corner of the configuration space $[-1,1]^N$). However, they are
the simplest possible in absence of retrieval states. We will adopt
them for this reason, but it should be kept in mind that they may not
be the most appropriate ones in general.

\begin{figure}[!]
\centering\includegraphics[width=7cm]{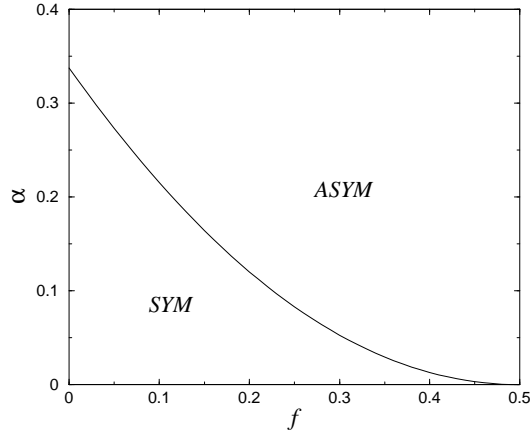}
\caption{\label{critl}Critical line separating the asymmetric,
inefficient phase with $H>0$ from the symmetric one with $H=0$ in the
$(f,\alpha)$ plane. As $\alpha\downarrow\alpha_c(f)$,
$\chi\to\infty$.}
\end{figure}

After some algebra, the set of saddle point equations can be greatly
simplified, because, as in \cite{mcz}, when $\beta_1,\beta_2\to\infty$
the averages (\ref{ave1}) and (\ref{ave2}) can be explicitly performed
by steepest descent. The result for the relevant quantities is
\begin{eqnarray}
 P=1\\
Q=1-\sqrt{\frac{2}{\pi}}\frac{e^{-\frac{\lambda^2}{2}}}{\lambda}-
\l(1-\frac{1}{\lambda^2}\r)\erf
\frac{\lambda}{\sqrt{2}}\\
\frac{\alpha \chi}{1+\chi}=(1-f)\erf 
\frac{\lambda}{\sqrt{2}}-f\sqrt{\frac{2}{\pi}}\lambda
\label{chi}
\end{eqnarray}
with $\lambda=\sqrt{\alpha/[1+(1-f)Q+f]}$. The identity $P=1$ implies
that majority agents use only one of their strategies, i.e. that the
stationary state of a pure majority game is in pure strategies. We
define
\begin{equation}\label{crep}
c=(1-f)Q+f
\end{equation}
Evidently, $H$ can be expressed in terms of all saddle-point values
since $\mathcal{H}/N=H$. Using (\ref{mimma}) and taking the limit
$R\to -1$ (this is equivalent to taking the limit $\beta_1\to\beta_2$
followed by $\beta_2\to\infty$) one easily finds
\begin{equation}
H=\frac{1+c}{2(1+\chi)^2}
\label{pred}
\end{equation}
The existence of a transition at some critical value of $\alpha$ is
determined by the divergence of $\chi$ (which means that $H$ becomes
$0$). From (\ref{chi}) we find for $\alpha_c$ the following
expression:
\begin{equation}\label{alfaccc}
\alpha_c(f)=(1-f)\erf(x)-\frac{2fx}{\sqrt{\pi}}
\end{equation}
where $x$ is the solution of
\begin{equation}\label{yaaa}
2-(1-f)\erf(x)-\frac{1-f}{x\sqrt{\pi}}e^{-x^2}+\frac{f}{x\sqrt{\pi}}=0
\end{equation}
Solving the above equations numerically for different $f$ one obtains
a very good agreement with the behavior of $H$ (see
Fig.~\ref{figs2H}). The critical line $\alpha_c$ calculated from
(\ref{alfaccc},\ref{yaaa}) is instead displayed in Fig.~\ref{critl}.
It should be mentioned that an approximate expression for $\sigma^2$
can also be obtained, $\sigma^2\simeq H+(1-c)/2$, but it is not as
accurate as the one for $H$. A better estimate of $\sigma^2$ is
obtained by solving the dynamics. As a last remark, let us notice that
for a pure majority game one gets, from (\ref{majo1}--\ref{majo3}) and
from the fact that $\avg{zm_2}_2=\sqrt{2/\pi}$,
\begin{equation}\label{hoppp}
\chi_2=\frac{1}{1+\sqrt{\alpha\pi}}\quad{\rm and}\quad
H=(1-\chi_2)^{-2}
\end{equation}
The expression for $\chi_2$ is identical to that of the Hopfield model
at zero temperature.

%%%%%%%%%%%%%%%%%%%%%%%%%%%%%%%%%%%%%%%%%%%%%%%%%%%%%%%%%%%%%%%%%

\section{Dynamics}

Let us turn our attention to the dynamics. For simplicity, we
concentrate on the `batch' case \cite{hc}, which is obtained by
averaging (\ref{dyna}) over the $\mu$'s and re-scaling time. This
amounts to considering the case in which performance updates are made
after many ($\Or(P)$) iterations rather than at end of every
round. This approximation has already proved to be an extremely good
one for minority games. One arrives at
\begin{equation}\label{batch}
y_i(t+1)-y_i(t)=\epsilon_i h_i+\epsilon_i\sum_{j=1,N}J_{ij}s_j(t)
\end{equation}
where $h_i=(2/\sqrt{N})\sum_{\mu=1,P}\xi_i^\mu\Omega^\mu$ and
$J_{ij}=(2/N)\sum_{\mu=1,P}\xi_i^\mu\xi_j^\mu$. The dynamical
partition function of (\ref{batch}) reads
\begin{eqnarray}
\fl
Z[\boldsymbol{\psi}]=\pathavg{e^{\ii\sum_{it}y_i(t)\psi_i(t)}}\nonumber\\
\fl =\int e^{\ii
\sum_{it}\widehat{y}_i(t)\l[y_i(t+1)-y_i(t)-\epsilon_i
h_i-\epsilon_i\sum_j
J_{ij}s_j(t)-\theta_i(t)\r]+y_i(t)\psi_i(t)}p(\boldsymbol{y}(0))
D(\boldsymbol{y},\boldsymbol{\widehat{y}})
\end{eqnarray}
where $D(\boldsymbol{y},\boldsymbol{\widehat{y}})=
\prod_{it}[dy_i(t)d\widehat{y}_i(t)/(2\pi)]$ and $\theta_i$ is a
time-dependent external field. In principle, disorder-averaged
correlation and response functions can be calculated exactly at all
times by taking appropriate derivatives of the disorder-averaged $Z$,
i.e.
\begin{eqnarray}
\fl \ovl{Z[\boldsymbol{\psi}]}=\int e^{\ii
\sum_{it}\widehat{y}_i(t)\l[y_i(t+1)-y_i(t)-\theta_i(t)\r]+
y_i(t)\psi_i(t)+N F(\boldsymbol{\widehat{y}})} p(\boldsymbol{y}(0))
D(\boldsymbol{y},\boldsymbol{\widehat{y}})\\
F(\boldsymbol{\widehat{y}})=\frac{1}{N}\log\ovl{\l[e^{-\ii\sum_{it}
\widehat{y}_i(t)\epsilon_i\l[h_i+\sum_j J_{ij}s_j(t)\r]}\r]}
\end{eqnarray}
with respect to the fields $\psi_i$ and $\theta_i$. We shall however
be interested in the stationary state only. As usual, evaluation of
$\ovl{Z}$ leads to an effective (non-Markovian) process that provides
an equivalent description of the original (Markovian) multi-agent
process (\ref{batch}). Such a calculation is in this case rather
similar to that done for the pure batch minority game in \cite{hc},
and is sketched in the Appendix. The main difference is that here we
obtain {\em two} effective processes, describing trend-followers and
fundamentalists respectively. These are given by
\begin{equation}\label{esap}
y(t+1)-y(t)=\alpha\epsilon\sum_{t'}[(\mathsf{I}+
\mathsf{G})^{-1}]_{tt'}s(t')+\theta(t)+ \sqrt{\alpha}z(t)
\end{equation}
where $\epsilon=1$ (resp. $-1$) for the majority (resp. minority)
part, and $z(t)$ is a zero-average Gaussian random variable with
temporal correlations
\begin{equation}
\avg{z(t)z(t')}=[(\mathsf{I}+\mathsf{G})^{-1}
(\mathsf{E}+\mathsf{C})(\mathsf{I}+ \mathsf{G}^T)^{-1}]_{tt'}
\end{equation}
$\mathsf{I}$ stands for the identity matrix while $\mathsf{E}$ denotes
the matrix with all elements equal to one. $\mathsf{C}$ has elements
$C_{tt'}=\avg{s(t)s(t')}$. $\mathsf{G}$, instead (see Appendix for
details), is the sum of two contributions:
\begin{equation}
\mathsf{G}=(1-f)\mathsf{G}_1-f\mathsf{G}_2
\end{equation}
$\mathsf{G}_1$ (resp. $\mathsf{G}_2$) has elements $\avg{\partial
s(t)/\partial\theta(t')}_{-1}$ (resp. $\avg{\partial
s(t)/\partial\theta(t')}_{+1}$) where the subscript means average over
the process (\ref{esap}) with $\epsilon=-1$ (resp. $+1$). When
$N\to\infty$, $C_{tt'}$ can be identified with the disorder- and
agent-averaged autocorrelation function of (\ref{batch}), while the
two components of $G_{tt'}$ become identical to the disorder- and
agent-averaged response functions of minority and majority agents,
respectively. 

Ergodic stationary states can be studied under the following
assumptions:
\begin{itemize}
\item Time-translation invariance (TTI): $\cases{
\lim_{t\to\infty}C_{t+\tau,t}=C(\tau)\\
\lim_{t\to\infty}G_{t+\tau,t}=G(\tau)}$;
\item Finite integrated `response' (FIR):
$\lim_{t\rightarrow\infty}\sum_{t'\leq t}G_{tt'}=\chi<\infty$;
\item Weak long-term memory (WLTM):
$\lim_{t\to\infty}G(t,t')=0\quad\forall t'$ finite.
\end{itemize}
The breakdown of any of these signals the breakdown of ergodicity. To
be more clear, we remark that the `integrated response' $\chi$ defined
in FIR has two components, i.e., with obvious notation,
\begin{equation}\label{chiii}
\chi=(1-f)\chi_1-f\chi_2
\end{equation}
and can be negative. $\chi_1$ and $\chi_2$ are the actual
susceptibilities of minority and majority agents, respectively. With
FIR, we will require that {\em both} $\chi_1$ and $\chi_2$ are finite.

As in the minority game, for individual agents there are two
possibilities: either $y_i(t)/t\to{\rm constant}\neq 0$ as
$t\to\infty$, in which case they use only one of their strategies
asymptotically (we call these agents ``frozen''); or $y_i(t)/t\to 0$
as $t\to\infty$, in which case they keep flipping between their
strategies even in the long run (we call these agents
``fickle''). Macroscopic quantities can be obtained by separating the
contributions of frozen and fickle agents.

\begin{figure}[t]
\begin{center}
\subfigure{\scalebox{.39}{\includegraphics{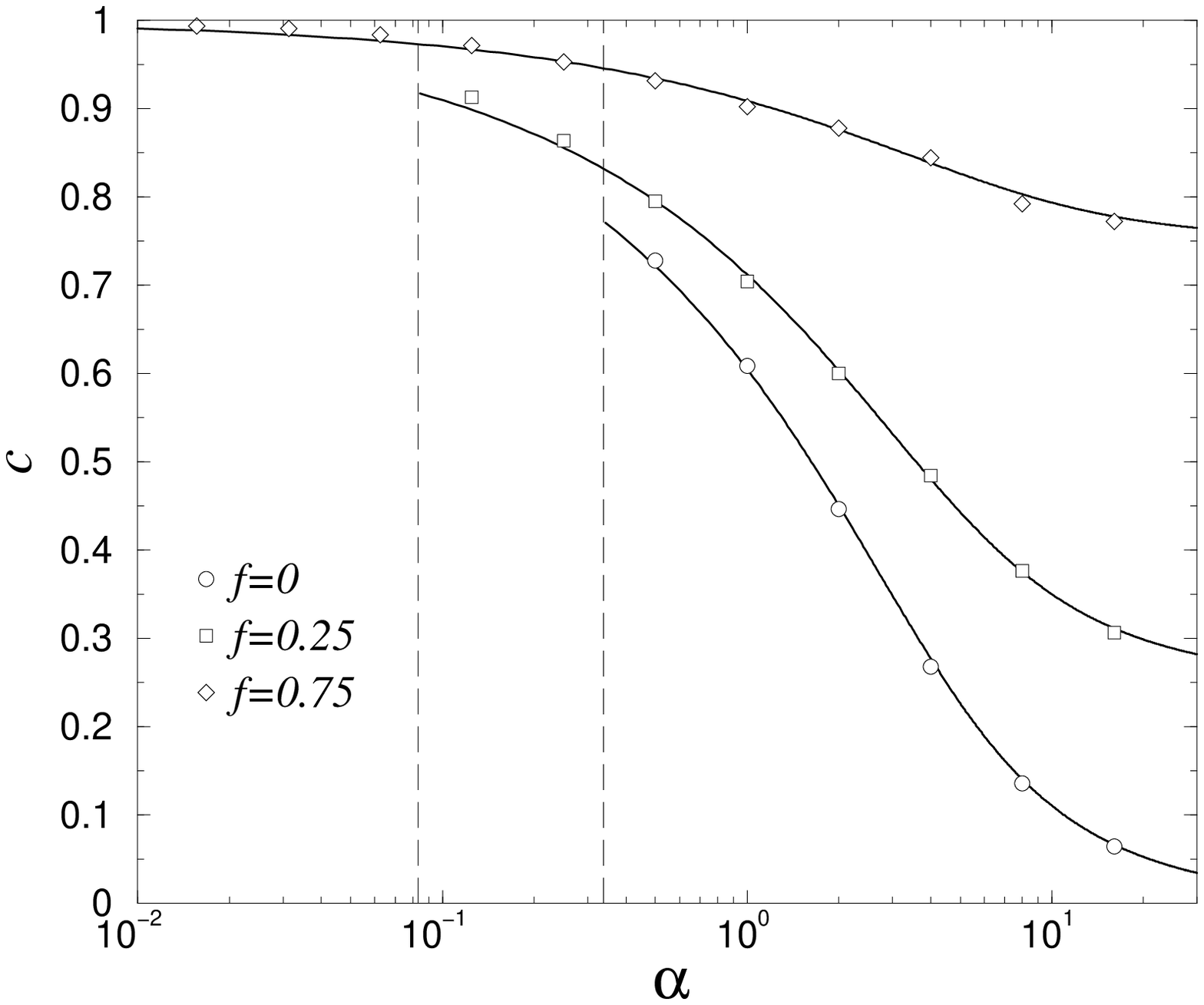}}}
\subfigure{\scalebox{.39}{\includegraphics{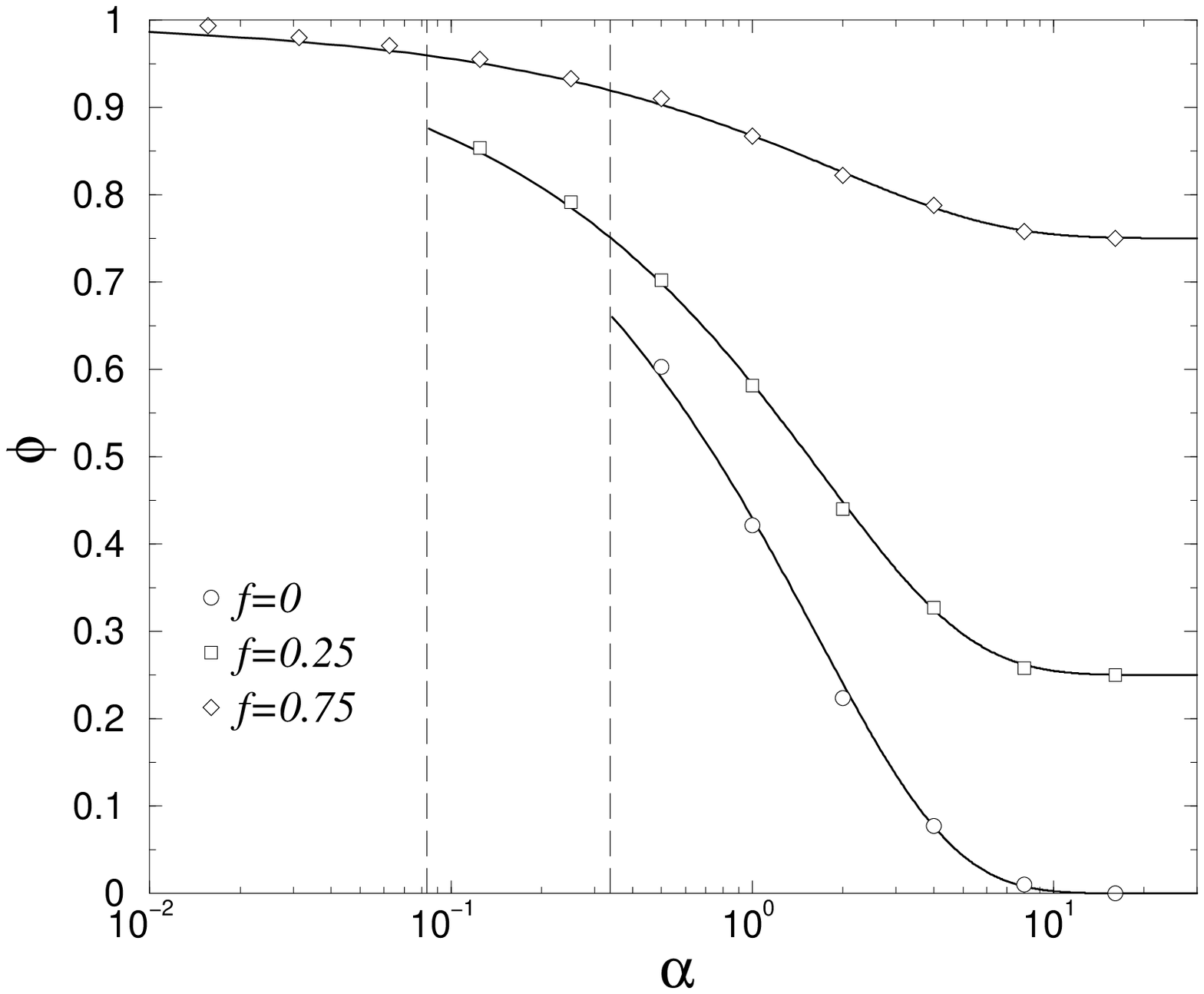}}}
\caption{\label{cphi}Persistent autocorrelation $c$ (left) and
fraction of frozen agents $\phi$ (right) for various $f$. Lines
correspond to the analytic solutions from  (\ref{c}) and
(\ref{phi}), markers are the results from numerical
simulations. Vertical lines give, for $f<1/2$, the position of the
critical points $\alpha_c$ below which the stationary state (hence $c$
and $\phi$) depends on initial conditions.}
\end{center}
\end{figure}

Defining $\widetilde{y}=\lim_{t\to\infty}y(t)/t$,
$s=\lim_{\tau\to\infty}(1/\tau)\sum_{t\leq\tau}\sign[y(t)/t]$ and
$z=\lim_{\tau\to\infty}(1/\tau)\sum_{t\leq\tau}z(t)$, one has that
\begin{equation}
\widetilde{y}=\frac{\alpha \epsilon s}{1+\chi}+\sqrt{\alpha}z +\theta=
\sqrt{\alpha}\epsilon\gamma s+\sqrt{\alpha}z+\theta
\end{equation}
Let us assume that $\gamma>0$ (this assumption is verified a
posteriori). For minority game players ($\epsilon=-1$), we have a
frozen agent (with $s=\sign(\widetilde{y})$) if $|z|>\gamma$ and a
fickle or non-frozen agent (with $s=z/\gamma$) if $|z|<\gamma$
\cite{hc}. In the majority part, all agents turn out to be frozen. In
particular, for $z>\gamma$ agents freeze at $s=1$, for $z<-\gamma$
they freeze at $s=-1$, while for $|z|<\gamma$ they can freeze at
either values of $s$. It follows that the average autocorrelation
$c=\lim_{\tau\to\infty}\frac{1}{\tau}\sum_{t\leq\tau}C(t)$ is given
by, separating the contributions of minority agents from majority
agents ($\avg{~}=$ average over Gaussian r.v. $z$ with variance
$\avg{z^2}=\lim_{\tau,\tau'\to\infty}(\tau\tau')^{-1}
\sum_{t\leq\tau,t'\leq\tau'} [(\mathsf{I}+
\mathsf{G})^{-1}(\mathsf{E}+\mathsf{C})(\mathsf{I}+
\mathsf{G}^T)^{-1}]_{tt'} =(1+\chi)^{-2}(1+c)$):
\begin{eqnarray}\label{c}
c&=(1-f)\l[\avg{\theta(|z|-\gamma)}+
\avg{\theta(\gamma-|z|)\frac{z}{\gamma}}\r]+f\nonumber\\
&=(1-f)\l[1-\erf\frac{\lambda}{\sqrt{2}}+\frac{1}{\lambda^2}
\l(\erf\frac{\lambda}{\sqrt{2}}-\lambda\sqrt{\frac{2}{\pi}}
e^{-\lambda^2/2}\r)\r]+f
\end{eqnarray}
where $\lambda=\sqrt{\frac{\alpha}{1+c}}$. This agrees with the
replica result (\ref{crep}). For the fraction $\phi$ of frozen agents
one obtains
\begin{equation}\label{phi}
\phi=(1-f)\avg{\theta(|z|-\gamma)}+f=
1-(1-f)\erf\frac{\lambda}{\sqrt{2}}
\end{equation}
In Fig.~\ref{cphi} analytical results for $c$ and $\phi$ are compared
with simulations.

The `susceptibility' (\ref{chiii}) can instead be calculated from the
formula
\begin{equation}
\chi=(1-f)\frac{\avg{sz}_{\rm min}}{\sqrt{\alpha}\avg{z^2}}-
f\frac{\avg{sz}_{\rm maj}}{\sqrt{\alpha}\avg{z^2}}
\end{equation}
which follows directly from the fact that response functions for
minority (resp. majority) agents can be obtained as\footnote{This is
due to the fact that the noise term and the external field enter
(\ref{esap}) in the same way, apart from the $\sqrt{\alpha}$ factor.}
$\alpha^{-1/2}\avg{\partial\sign[y(t)]/\partial z(t')}_{-1}$
(resp. $\alpha^{-1/2}\avg{\partial\sign[y(t)]/\partial z(t')}_{+1}$),
after an integration by parts and a time average \cite{hc}. The
minority part is as usual given by
\begin{equation}
\avg{sz}_{\rm min}=\avg{\theta(|z|-\gamma)|z|}+
\avg{\theta(\gamma-|z|)\frac{z^2}{\gamma}}=
\frac{1+c}{\sqrt{\alpha}(1+\chi)}~\erf\frac{\lambda}{\sqrt{2}}
\end{equation}
To calculate the majority part, one must fix the value of the product
$sz$ for $-\gamma\leq z\leq\gamma$, where $s$ can be either $+1$ or
$-1$ (for $|z|>\gamma$ one has $sz=|z|$ in any case). In principle,
there are several possibilities. If one makes the `natural' choice
$s=\sign(z)$, then
\begin{equation}\label{sz1}
\avg{sz}_{\rm maj}=\avg{|z|}=\sqrt{\frac{2}{\pi}}
\sqrt{\frac{1+c}{(1+\chi)^2}}
\end{equation}
This leads to
\begin{equation}\label{chi1}
\frac{\alpha\chi}{1+\chi}=(1-f)\erf\frac{\lambda}{\sqrt{2}}-
f\lambda\sqrt{\frac{2}{\pi}}
\end{equation}
$\chi$ diverges (hence FIR is violated and ergodicity is broken) when
the fraction $\ovl{\phi}=1-\phi$ of fickle agents satisfies
$\ovl{\phi}=\alpha+f\lambda\sqrt{2/\pi}$ or, equivalently, at the
critical values of $\alpha$ given by the equation
\begin{equation}\label{alfacr1}
\alpha_c(f)=(1-f)\erf(x)-\frac{2fx}{\sqrt{\pi}}
\end{equation}
where $x$ is the solution of
\begin{equation}\label{cx}
2-(1-f)\erf(x)-\frac{1-f}{x\sqrt{\pi}}e^{-x^2}+\frac{f}{x\sqrt{\pi}}=0
\end{equation}
(\ref{chi1}--\ref{cx}) are in full agreement with the replica results
of Sec. 3.

Another possibility is to calculate $\avg{sz}_{\rm maj}$ without
making any special assumption on $s$ for $-\gamma\leq z\leq\gamma$.
This brings us to a situation where (\ref{sz1}--\ref{alfacr1}) are
substituted respectively by
\begin{eqnarray}
\avg{sz}_{\rm maj}&=\avg{\theta(z+\gamma)z}-\avg{\theta(\gamma-z)
z}=e^{-\lambda^2/2}\sqrt{\frac{2}{\pi}}\sqrt{\frac{1+c}{(1+\chi)^2}}\\
\frac{\alpha\chi}{1+\chi}&=(1-f)\erf\frac{\lambda}{\sqrt{2}}-
f\lambda\sqrt{\frac{2}{\pi}}e^{-\lambda^2/2}\\
\alpha_c(f)&=(1-f)\erf(x)-\frac{2fx}{\sqrt{\pi}}e^{-x^2}
\end{eqnarray}
where $x$ now solves
\begin{equation}
2-(1-f)\erf(x)-\frac{1-2f}{x\sqrt{\pi}}e^{-x^2}=0
\end{equation}
The value of $\ovl{\phi}$ at which $\chi$ diverges is now
$\ovl{\phi}=\alpha+f\lambda e^{-\lambda^2/2}\sqrt{2/\pi}$. Notice that
the extra exponential factor one obtains in this way does not change
numerical results for $\alpha_c$ significantly (the solution of
(\ref{cx}) is in fact $\lesssim 0.3$, so $e^{-x^2}$ is always close to
$1$). Notice also that for a purely majority game (recalling that
$\chi_2=-\chi$) one gets for the susceptibility
\begin{equation}
\chi_2=\frac{e^{-\alpha/4}/\sqrt{\alpha\pi}}
{1+e^{-\alpha/4}/\sqrt{\alpha\pi}}
\end{equation}
instead of the Hopfield-like formula (\ref{hoppp}). In both cases,
$\chi_2\to\infty$ when $\alpha\downarrow 0$.

For the stationary volatility, which reads \cite{hc}
\begin{equation}
\sigma^2=\frac{1}{2}\lim_{t\to\infty} [(\mathsf{I}+\mathsf{G})^{-1}
(\mathsf{E}+\mathsf{C})(\mathsf{I}+ \mathsf{G}^T)^{-1}]_{tt}
\end{equation}
one can use the approximate method of \cite{hc} to derive an
expression in terms of the persistent parameters $\chi$ and $\phi$,
which holds for $\alpha>\alpha_c$:
\begin{equation}
\sigma^2=\frac{1+\phi}{2(1+\chi)^2}+\frac{1}{2}(1-\phi)
\end{equation}
Solving for $\chi$, $\phi$ and $c$ for different $f$ and substituting
one obtains the volatility branches displayed in Fig.~\ref{figs2H},
which are again in excellent agreement with simulations.

%%%%%%%%%%%%%%%%%%%%%%%%%%%%%%%%%%%%%%%%%%%%%%%%%%%%%%%%%%%%%%%%%

\section{Summary and outlook}

To summarize, we have studied the mixed majority-minority game with
random external information. Neglecting `retrieval' (i.e. the
possibility that trend-followers flock in presence of a particular
information pattern), we have first calculated the stationary state of
the dynamics from a static approximation via a negative-replica
theory. Then we have solved the dynamics using generating functional
methods. The two approaches match nicely and agree with numerical
results for the macroscopic observables $\sigma^2$ and $H$ in a
satisfactory way. This suggests that retrieval does not affect such
quantities significantly. Our results also indicate that when
fundamentalists outnumber trend-followers, the macroscopic behavior of
the system (`phase transition' with ergodicity breaking from an
inefficient phase at high $\alpha$ to an efficient one at low
$\alpha$) can be explained by the onset of anomalous response, that is
by a divergence of the integrated response, as in the pure minority
game. We have calculated the line of critical points in the
$(f,\alpha)$ plane showing that the inefficient phase gets larger as
$f$ increases. When trend-followers dominate, instead, the system is
always inefficient and low volatility states disappear. As a
byproduct, we have provided an approximate static and dynamical
solution of the majority game. A greater effort is nevertheless needed
in order to incorporate the possibility of `herding' in both the
replica theory and the path-integral solution. We expect retrieval
states to exist at low $\alpha$ for any $f>0$. While such states
shouldn't affect global time-averaged properties (i.e. $\sigma^2$ and
$H$) significantly, they are likely to play a most crucial role in
such phenomena as ``bubbles'', that in our setting can be seen as
localized in time. It is also likely that RSB occurs at very low
$\alpha$ for any $f>0$, in pretty much the same way as RSB occurs for
any non-zero market impact in the pure minority game
\cite{mcz,dema,hd}.

Let us finally remark two aspects of the present model that can be
criticized and hence improved. In first place, all players can in
principle win at the same time (i.e. the available resources are
infinite), which is a clearly unrealistic situation (albeit extremely
unlikely in our disordered setup with $N\to\infty$). Secondly, in a
market a large buy rush today is justified by the belief that {\em
tomorrow} the price will rise again so that for instance one will be
able to sell at a higher price. So in a majority game it would perhaps
be more correct to measure the effectiveness of a trading decision
made today by what the payoff will have been tomorrow
\cite{gb2,sornette}. In other words, a player making a trading
decision $a_i(n)$ at round $n$ should receive a payoff
$u_i(n+1)=a_i(t)A(n+1)$ at round $n+1$. Instead, in our model, his
payoff is $u_i(n)=a_i(n)A(n)$. In spite of these limitations, we see
that our model does indeed capture some of the features one expects to
find in markets where fundamentalists and trend-followers
compete. Also, we believe that some of the issues listed above,
starting with retrieval, can be taken into account, possibly with
modest modifications. In our view, a possibly more interesting
generalization would consist in allowing the $\epsilon_i$'s to be
dynamical variables, in order to give agents the possibility to change
their character. Some work along these lines is currently in progress.

%%%%%%%%%%%%%%%%%%%%%%%%%%%%%%%%%%%%%%%%%%%%%%%%%%%%%%%%%%%%

\ack We are grateful to A Cavagna, M Marsili, G Parisi and F
Ricci-Tersenghi for useful discussions and suggestions, and in
particular to M Marsili for disclosing some results of \cite{matmaj}
prior to publication. We would also like to thank ACC Coolen for
introducing us to dynamical methods.

%%%%%%%%%%%%%%%%%%%%%%%%%%%%%%%%%%%%%%%%%%%%%%%%%%%%%%%%%%%%%%%%%

%\appendix

\section*{Appendix: Generating functional analysis}

The disorder average is as usual expected to generate two-time
player-averaged functions of the $s_i$ and $\widehat{y}_i$ variables
only. We focus on
\begin{eqnarray}
L_{tt'}&=\frac{1}{N}\sum_{i=1,N}\widehat{y}_i(t)\widehat{y}_i(t')\\
Q_{tt'}&=\frac{1}{N}\sum_{i=1,N}s_i(t)s_i(t')\\
K_{tt'}&=-\frac{1}{N}\sum_{i=1,N}\epsilon_i
s_i(t)\widehat{y}_i(t')
\end{eqnarray}
The matrix $\mathsf{K}$ can be seen as formed by two components, for
minority and majority agents, respectively:
\begin{equation}
\mathsf{K}=(1-f)\mathsf{K}_1-f\mathsf{K}_2
\end{equation}
Forcing the above definitions inside $\ovl{Z}$ via $\delta$-functions
with the proper $N$-scaling and assuming that
$p(\boldsymbol{y}(0))=\prod_{i=1,N}p(y_i(0))$, we find (with the
shorthand $D(\mathsf{X},\mathsf{\widehat{X}})=
\prod_{tt'}dX_{tt'}d\widehat{X}_{tt'}/(2\pi)$)
\begin{equation}
\ovl{Z[\boldsymbol{\psi}]}=\int e^{N(\Psi+\Omega+\Phi)}
D(\mathsf{Q},\mathsf{\widehat{Q}})
D(\mathsf{L},\mathsf{\widehat{L}})
D(\mathsf{K},\mathsf{\widehat{K}})
\end{equation}
where
$\Psi(\mathsf{Q},\mathsf{\widehat{Q}},\mathsf{L},\mathsf{\widehat{L}},
\mathsf{K},\mathsf{\widehat{K}})=
\ii\Tr[\mathsf{\widehat{Q}}^T\mathsf{Q}+
\mathsf{\widehat{L}}^T\mathsf{L}+\mathsf{\widehat{K}}^T\mathsf{K}]$,
\begin{eqnarray}
\fl
\Omega(\mathsf{\widehat{Q}},\mathsf{\widehat{L}},\mathsf{\widehat{K}})=
\frac{1}{N}\sum_{i=1,N}\log\int D(y,\widehat{y}) p(y(0))~e^{\ii
\sum_{t} \widehat{y}(t) [y(t+1)-y(t)-\theta_i(t)]
+y(t)\psi_i(t)}\times\nonumber\\\times e^{-\ii
\sum_{tt'}[s(t)\widehat{Q}_{tt'}s(t')+\widehat{y}(t) \widehat{L}_{tt'}
\widehat{y}(t')-\epsilon_i s(t)\widehat{K}_{tt'}\widehat{y}(t')]}
\end{eqnarray}
and
$\Phi(\mathsf{Q},\mathsf{L},\mathsf{K})=F(\boldsymbol{\widehat{y}})$.
To calculate the latter, it suffices to make use of the definitions of
$h_i$ and $J_{ij}$ and to introduce, via $\delta$-functions, the
parameters
\begin{equation}
x_t^\mu=\sqrt{\frac{2}{N}}\sum_{i=1,N}s_i(t)\xi_i^\mu\quad {\rm and}
\quad
w_t^\mu=-\sqrt{\frac{2}{N}}\sum_{i=1,N}\epsilon_i\widehat{y}_i(t)
\xi_i^\mu
\end{equation}
It turns out that the relevant term for the disorder average is
\begin{eqnarray}
\fl \ovl{e^{ \ii\sqrt{2} \sum_{t\mu} w_t^\mu\Omega^\mu -\ii
\sqrt{\frac{2}{N}} \sum_{i\mu} \xi_i^\mu \sum_t [\widehat{x}_t^\mu
s_i(t)-\widehat{y}_i(t)\epsilon_i \widehat{w}_t^\mu]}}=\nonumber\\
\hspace{2cm}
=e^{-\frac{1}{2}\sum_{tt'\mu}(w_t^\mu w_{t'}^\mu + \widehat{w}_t^\mu
L_{tt'} \widehat{w}_{t'}^\mu+\widehat{x}_t^\mu Q_{tt'}
\widehat{x}_{t'}^\mu+2\widehat{x}_t^\mu K_{tt'} \widehat{w}_{t'}^\mu)}
\end{eqnarray}
so that finally one has (with $D(z,\widehat{z})=\prod_t
dx_td\widehat{x}_t/(2\pi)$)
\begin{eqnarray}
\fl\Phi(\mathsf{Q},\mathsf{L},\mathsf{K})=\alpha\log\int
D(x,\widehat{x}) D(w,\widehat{w})\times\nonumber\\ \times e^{\ii\sum_t
(x_t\widehat{x}_t+w_t\widehat{w}_t+x_t w_t)-\frac{1}{2}\sum_{tt'}[w_t
w_{t'} + \widehat{w}_t L_{tt'} \widehat{w}_{t'}^\mu+\widehat{x}_t
C_{tt'} \widehat{x}_{t'}+2\widehat{x}_t K_{tt'} \widehat{w}_{t'}]}
\end{eqnarray}
where all integrals are from $-\infty$ to $+\infty$.  

In the limit $N\to\infty$ the dominant contribution to
$\ovl{Z[\boldsymbol{\psi}]}$ comes from the saddle point described by
the equations
\begin{eqnarray} 
\fl \ii \widehat{Q}_{tt'}=-\partial_{Q_{tt'}}\Phi \qquad \ii
\widehat{L}_{tt'}=-\partial_{L_{tt'}}\Phi \qquad \ii
\widehat{K}_{tt'}=-\partial_{K_{tt'}}\Phi\\ \fl
Q_{tt'}=\stavg{s(t)s(t')} \hspace{0.5cm}
L_{tt'}=\stavg{\widehat{y}(t)\widehat{y}(t')} \hspace{0.5cm}
K_{tt'}=-\stavg{\epsilon_i s(t)\widehat{y}(t')}
\end{eqnarray}
where
\begin{equation}
\stavg{h(s,y,\widehat{y})}=\frac{1}{N}\sum_{i=1,N}\frac{\int
h(s,y,\widehat{y})
M_i^{\epsilon_i}(s,y,\widehat{y})D(y,\widehat{y})}{\int
M_i^{\epsilon_i}(s,y,\widehat{y})D(y,\widehat{y})}
\end{equation}
with
\begin{eqnarray}
\fl M_i^{\epsilon_i}(s,y,\widehat{y})= p(y(0))~e^{\ii \sum_{t}
\widehat{y}(t) [y(t+1)-y(t)-\theta_i(t)]
+y(t)\psi_i(t)}\times\nonumber\\ \times e^{-\ii
\sum_{tt'}[s(t)\widehat{Q}_{tt'}s(t')+\widehat{y}(t) \widehat{L}_{tt'}
\widehat{y}(t')-\epsilon_i s(t)\widehat{K}_{tt'}\widehat{y}(t')]}
\end{eqnarray}
It can be checked by a direct calculation (e.g. following \cite{hc})
that, at the relevant saddle point,
\begin{equation}
Q_{tt'}=C_{tt'}\equiv\frac{1}{N}\sum_{i=1,N}
\ovl{\pathavg{s_i(t)s_i(t')}}\quad{\rm and}~~
L_{tt'}=0
\end{equation}
As for $K_{tt'}$, one can define $-\ii\mathsf{K}=\mathsf{G}$ and see,
for instance by taking the derivative of $\ovl{\pathavg{s_i(t)}}$ with
respect to $\theta_i(t')$, that
\begin{equation}
\mathsf{G}=(1-f)\mathsf{G}_1-f\mathsf{G}_2
\end{equation}
where $\mathsf{G}_1$ is the response function of minority agents, with
elements
\begin{equation}
G_{tt'}^{(1)}=\frac{1}{N_1}\sum_{i\in
N_1}\frac{\partial}{\partial\theta_i(t')} \ovl{\pathavg{s_i(t)}}
\end{equation}
and similarly $\mathsf{G}_2$ is the response function of majority
agents.

Setting the generating field $\psi_i$ to zero and assuming that
$\theta_i(t)=\theta(t)$, we can now treat minority agents
($\epsilon_i=-1$) and majority agents ($\epsilon_i=1$) separately.
We get, for $\Omega$:
\begin{eqnarray}
\fl \Omega^\epsilon=\log\int e^{\ii \sum_{t} \widehat{y}(t)
[y(t+1)-y(t)-\theta(t)]}\times\nonumber\\ \times e^{-\ii
\sum_{tt'}[s(t)\widehat{C}_{tt'}s(t')+\widehat{y}(t) \widehat{L}_{tt'}
\widehat{y}(t')-\epsilon s(t)\widehat{K}_{tt'}\widehat{y}(t')]}p(y(0))
D(y,\widehat{y})
\end{eqnarray}
where we set $\mathsf{\widehat{Q}}=\mathsf{\widehat{C}}$; the measure
$M_i^{\epsilon_i}$ instead becomes
\begin{eqnarray}
\fl M^\epsilon(s,y,\widehat{y})=p(y(0))
e^{-\ii\sum_{tt'}s(t)\widehat{C}_{tt'}s(t')}\times\nonumber\\ \times
e^{-\ii\sum_{tt'}
\widehat{y}(t)\widehat{L}_{tt'}\widehat{y}(t')+\ii\sum_t
\widehat{y}(t)
[y(t+1)-y(t)-\theta(t)+\epsilon\sum_{t'}\widehat{K}^T_{tt'}s(t)]}
\end{eqnarray}
$M^1$ and $M^{-1}$ represent majority and minority agents,
respectively. The saddle-point equations for $\mathsf{\widehat{C}}$,
$\mathsf{\widehat{L}}$ and $\mathsf{\widehat{K}}$ are identical to
those found for the pure batch minority game \cite{hc}. It results
that
\begin{eqnarray}
\widehat{C}_{tt'}=0\nonumber\\
\widehat{K}^T_{tt'}=-\alpha[(\mathsf{I}-\ii
\mathsf{K})^{-1}]_{tt'}\\
\widehat{L}_{tt'}=-\frac{1}{2}\ii\alpha[(\mathsf{I}-\ii
\mathsf{K})^{-1} (\mathsf{E}+\mathsf{C})(\mathsf{I}-\ii
\mathsf{K}^T)^{-1}]_{tt'}\nonumber
\end{eqnarray}
where $I_{tt'}=\delta_{tt'}$ and $E_{tt'}=1$. Substituting these into
$M^\epsilon$ one obtains
\begin{eqnarray}
\fl M^\epsilon(s,y,\widehat{y})=p(y(0))
e^{-\frac{1}{2}\alpha\sum_{tt'}\widehat{y}(t)[(\mathsf{I}-\ii
\mathsf{K})^{-1} (\mathsf{E}+\mathsf{C})(\mathsf{I}-\ii
\mathsf{K}^T)^{-1}]_{tt'}\widehat{y}(t')}\times\nonumber\\\times
e^{\ii\sum_t\widehat{y}(t)
[y(t+1)-y(t)-\theta(t)-\alpha\epsilon\sum_{t'}[(\mathsf{I}-\ii
\mathsf{K})^{-1}]_{tt'} s(t)]}
\end{eqnarray}
Recalling that $\mathsf{K}=\ii\mathsf{G}$, it turns out that the
disorder-averaged correlation and response functions for minority and
majority agents are obtained as averages over the colored effective
stochastic processes (\ref{esap}) with $\epsilon=-1$ and $\epsilon=1$,
respectively.

%%%%%%%%%%%%%%%%%%%%%%%%%%%%%%%%%%%%%%%%%%%%%%%%%%%%%%%%%%%%%%%%%

\end{document}